\documentclass[a4paper,preprint,aps]{revtex4}

\usepackage{graphicx}
\usepackage{dcolumn}
\usepackage{bm}

\newcommand{\eqa}{\begin{equation}}
\newcommand{\eqz}{\end{equation}}
\newcommand{\eqma}{\begin{eqnarray}}
\newcommand{\eqmz}{\end{eqnarray}}

\begin{document}
\newcommand{\e}{{\em e}~}
\title{W4 thermochemistry of P$_2$ and P$_4$. 
Is the CODATA heat of formation of phosphorus atom correct?\footnote{Dedicated to Professor Peter Pulay on the occasion of his 65th birthday.}}
\author{Amir Karton}
\affiliation{Department of Organic Chemistry, 
Weizmann Institute of Science, 
IL-76100 Re\d{h}ovot, Israel. Email: {\tt comartin@weizmann.ac.il}}
\author{Jan M. L. Martin*}
\affiliation{Department of Organic Chemistry, 
Weizmann Institute of Science, 
IL-76100 Re\d{h}ovot, Israel.  Email: {\tt comartin@weizmann.ac.il}}

\date{
Received May 24, 2007; Accepted June 28, 2007. \\~{\em Mol. Phys.} (Peter Pulay festschrift) {\bf TMPH-2007-0140}}

\begin{abstract}
The high-accuracy W4 computational thermochemistry protocol, and several post-W4 methods, have been applied to the P$_2$ and P$_4$ molecules. Contrary to previous studies, we find the experimental thermochemistry to be fundamentally sound. The reaction enthalpy for P$_4\rightarrow 2$P$_2$ has a very significant contribution from post-CCSD(T) correlation effects.
We derive a gas-phase heat of formation for the phosphorus atom of $\Delta H^\circ_{f,0}$[P(g)]=75.54$\pm$0.1 kcal/mol and $\Delta H^\circ_{f,298}$[P(g)]=75.74$\pm$0.1 kcal/mol, in the upper half of the CODATA uncertainty interval.
\end{abstract}

\maketitle

\section{Introduction}

The most important component of phosphorus vapor is tetrahedral P$_4$. It is also a building block of the most stable solid allotropes (white and red phosphorus), as well as an important component of the liquid phase\cite{Jones}.

P$_4$ is a challenging molecule to describe well in terms of  both 1-particle basis set and $n$-particle space ("electron correlation treatment"). Its peculiar bonding pattern has been studied in some detail by Ahlrichs et al.\cite{Ahlrichs}. Perhaps the most comprehensive {\em ab initio} study to date has been the work of Persson, Taylor, and Lee (PTL)\cite{PTL}, who studied the thermochemistry, geometry, and anharmonic force field of the molecule at the MP2 (second-order many-body perturbation theory) and CCSD(T) (coupled cluster with all singles and doubles plus quasiperturbative triples\cite{Rag89}) levels with basis sets of up to $spdfg$ quality. Their work buttressed the earlier contention of H\"aser and Treutler\cite{Has95} that the established bond distance\cite{Bra81} is considerably too long due to misassignment of the rotational fine structure of the Raman spectrum. Subsequent IR work\cite{Bou99} confirmed this conclusion.

In addition, the PTL anharmonic force field calculations showed implausibly large discrepancies with the experimental band origins, which were resolved by subsequent remeasurements\cite{Kor02}.

Finally, PTL addressed the endothermicity of the reaction P$_4\rightarrow 2$P$_2$, and found a significant discrepancy with experiment, a substantial part of it resulting from a very large core-valence contribution. Haworth and Bacskay (HB)\cite{Haw02}, in a later benchmark thermochemical study on a number of phosphorus compounds including P$_2$ and P$_4$, were able to rationalize the core-valence contribution as an artifact of the basis set used. However, despite their work being carried out near the CCSD(T) basis set
limit and including both core-valence and scalar relativistic corrections,
HB's best calculated total atomization energies (TAE$_0$ values) still imply a reaction energy for P$_4\rightarrow 2$P$_2$ that is almost 2 kcal/mol more endothermic than experiment\cite{Gurvich}.

This is not a purely academic issue. This selfsame reaction represents the weakest link in the determination of the CODATA\cite{codata} gas-phase heat of formation of the phosphorus atom. (The other two links, namely the heat of vaporization of P$_4$ and the dissociation energy of P$_2$, are fairly precisely known\cite{codata,Hub79}.) Said heat of formation enters the equation whenever one calculates the heat of formation of any phosphorus compound --- be it by {\em ab initio}, semiempirical, or density functional techniques.

Very recently, we developed a next-generation, fully {\em ab initio}, computational thermochemistry method known as W4 theory\cite{w4}, that was shown to be able to reproduce the best ATcT (active thermochemical tables\cite{Branko1,Branko2,Branko3}) atomization energies for several dozen small molecules with an RMS deviation of 0.08 kcal/mol (95\% confidence interval of 0.16 kcal/mol). Among other things, this method was applied effectively in a recent revision\cite{BeBAlSi} of the atomic heats of formation of boron, aluminum, and silicon. It would appear to be an appropriate tool for trying to resolve the P$_4$ issue. In the present paper, we will show that the experimental thermochemistry in general,
and the CODATA $\Delta H^\circ_{f,0}$[P(g)] value in particular, are in fact fundamentally correct.

\section{Computational methods}

All calculations reported in the present work were performed on the
Linux cluster of the Martin group.
All SCF, CCSD and CCSD(T) calculations were carried out using MOLPRO
2006.1\cite{molpro}. Some CCSDT calculations 
were carried out using the Austin-Mainz-Budapest version of the ACES II\cite{aces2de}  program system;
the remaining post-CCSD(T) calculations were carried out using an OpenMP
parallel version of K\'allay's general coupled cluster code
MRCC\cite{mrcc,KallayGauss,Bomble2005} interfaced to the Austin-Mainz-Budapest version of
ACES II\cite{aces2de}. SCF and correlated DBOCs were obtained using the relevant module\cite{ValeevSherrill} of the PSI3 open source quantum chemistry code\cite{psi3}.
All basis sets employed
belong to the
correlation consistent family of Dunning and
coworkers\cite{Dun89,Ken92,Wilson,pwCVnZ}. Convergence of iterative processes was accelerated by means of Pulay's direct inversion of the iterative subspace (DIIS) method\cite{DIIS}.

The computational protocols for W4 theory, for the lower-cost methods W2.2, W3.2, and W4lite, and for the post-W4 methods W4.2, W4.3, and W4.4 have been documented in great detail elsewhere\cite{w4,pCb}, and will not be repeated here. For the sake of clarity in the following discussion, we do note that the W4 energy consists of the following components (summarized in Table I): SCF, valence CCSD correlation, valence parenthetical triples (T), higher-order connected triples $T_3-(T)$, quasiperturbative connected quadruples (Q), higher-order connected quadruples $T_4-(Q)$, connected quintuples, CCSD(T) level inner-shell correlation, scalar relativistic effects (second-order Douglas-Kroll-Hess approximation\cite{DKH}), first-order spin-orbit coupling, and diagonal Born-Oppenheimer correction (DBOC). W4.2 and higher add a contribution for core-valence higher-order triples, while W4.3 uses more extended basis sets in calculating and extrapolating the post-CCSD(T) correlation effects (besides adding a connected sextuples term). W4.4 theory\cite{pCb} employs improved extrapolations, as well as core-valence parenthetical quadruples and a correlation correction to the DBOC.

\section{Results and discussion}

\subsection{At the CCSD(T) limit}

All relevant thermochemical data can be found in Table I. Convergence of various contributions as a function of basis set is summarized in Table II.

The valence-correlation-only CCSD(T)/cc-pV(Q+d)Z bond distance for P$_4$, 2.1987 \AA, is nearly 0.01 \AA\ longer than the all-electron CCSD(T)/cc-pwCVQZ bond distance, 2.1893 \AA. Combining this with the computed $r_0-r_e$ of PTL, 0.005 \AA, this implies $r_0$=2.194 \AA, in excellent agreement with the experimental value of Boudon et al.\cite{Bou99}, $r_0$=2.1958 \AA. (An earlier measurement\cite{Bra81} of 2.2228$\pm$0.0005 \AA\ was previously rejected by PTL on the basis of their calculations, which were subsequently confirmed by the Boudon et al. work.)

The valence-only and all-electron calculated bond distances of P$_2$ are 1.9019 and 1.8941 \AA, respectively. The latter value is in excellent agreement with the experimental one from Huber and Herzberg\cite{Hub79}, $r_e$=1.8934 \AA.

It was recently noted, in the context of a basis set convergence study on higher-order correlation effects\cite{pCb}, that the use of valence-only correlated reference geometries has insignificant consequences for first-row systems, but has more noticeable effects for second row compounds with strong bonds. In the present work, we find a small but non-negligible 0.05 kcal/mol for P$_2$ and a whopping 0.13 kcal/mol for P$_4$ --- the latter is not surprising in light of the hefty geometry change.

The zero-point vibrational energy of P$_4$, 3.91 kcal/mol, was derived from combining the anharmonicity constants of PTL with the revised experimental fundamentals of Ref.\cite{Kor02}. The one for P$_2$, 1.11 kcal/mol, derives from the spectroscopic constants given in Huber and Herzberg\cite{Hub79}.

At the W4 level, the valence CCSD(T) contributions to TAE$_0$ add up to 116.09 kcal/mol for P$_2$, and 289.32 kcal/mol for P$_4$. These values are lower by 0.6--0.7 and 1.1--1.5 kcal/mol,  respectively, than the corresponding limits of HB. (The ranges given refer to the spread among the various basis set combinations extrapolated from.) Of these respective discrepancies, 0.12 (P$_2$) and 0.24 kcal/mol (P$_4$) reflect the different definitions of the restricted open-shell CCSD(T) energy used for phosphorus atom: Watts-Gauss-Bartlett\cite{Wat93}, a.k.a. RHF-UCCSD(T), in the present work, and Knowles-Hampel-Werner\cite{Ham92}, a.k.a., RHF-RCCSD(T), in HB. The remainder may somewhat reflect the different reference geometries used (HB employed B3LYP/6-31G(2df) density functional calculations): stretching or compressing bonds causes a redistribution of binding energy between valence and core-valence contributions that is much more significant than the overall change in the sum of these parts. A more important component of the difference probably results from the fact that while HB extrapolated the total valence CCSD(T) energy, the present authors carried out separate extrapolations of the SCF, valence CCSD correlation, and (T) contributions, and in addition extrapolate singlet-and triplet-coupled pair correlation energies separately as advocated by Klopper\cite{Klopper}. Furthermore, we use more extended aug-cc-pV(5+d)Z and aug-cc-pV(6+d)Z basis sets. Using smaller basis sets with our procedure --- as is done in W2.2 and W3.2 theory --- changes the atomization energies of P$_2$ and P$_4$ by just -0.07 and +0.08 kcal/mol, respectively, suggesting that our W4-level extrapolation should be close to Nature. 

Likewise, our core-valence contributions differ somewhat from those of HB: for P$_4$ we obtain 1.76 kcal/mol extrapolated from CCSD(T)/aug-cc-pwCVTZ and aug-cc-pwCVQZ basis sets, while their result with the rather small cc-pCVTZ basis set is 2.2 kcal/mol without counterpoise correction, and 1.4 kcal/mol with counterpoise correction. HB's core-valence contributions for P$_2$ likewise bracket ours.

Our scalar relativistic contributions, -0.28 (P$_2$) and -0.72 (P$_4$) kcal/mol, are somewhat larger in absolute value than those of HB (-0.2 and -0.5 kcal/mol, respectively): while they employ first-order Darwin and mass-velocity corrections\cite{Cow76} from a CASPT2\cite{CASPT2} wave function with the G3large basis set\cite{G3}, we use the second-order Douglas-Kroll-Hess\cite{DKH} approximation from a CCSD(T) wave function with a relativistically contracted aug-cc-pVQZ basis set\cite{relcc}. The latter approximation should normally be the more rigorous one.

\subsection{Beyond the CCSD(T) limit}

We shall now consider post-CCSD(T) contributions.
The \%[(T)] diagnostic\cite{w4} for P$_2$, 8.4\%, indicates a system with moderate, bordering on severe, nondynamical correlation. (The largest $T_2$ amplitude in the CCSD calculation, 0.10 per degenerate component, is related to the $\pi\rightarrow\pi^*$ excitation. For additional diagnostics, see Table III.) As a result, error compensation between higher-order triples (-1.00 kcal/mol at the W4 level) and connected higher-order excitations (+1.46 kcal/mol connected quadruples, +0.10 kcal/mol connected quintuples) is less than perfect, and any CCSD(T)-based thermochemistry protocol would significantly underestimate $D_0$. In fact, the difference between W4 and the higher-level W4.3 protocol, 0.20 kcal/mol, is the largest we have thus far encountered. The difference breaks down as 0.06 kcal/mol higher-order triples, 0.06 kcal/mol connected quadruples, 0.03 kcal/mol connected quintuples, 0.03 kcal/mol from the newly introduced connected sextuples term, and 0.03 kcal/mol from higher-order triples in the inner shell correlation term. Applying our highest-level protocol, W4.4 theory, only causes an additional 0.02 kcal/mol change in $D_0$: decreases in the CCSD and (T) extrapolated contributions are compensated by a 0.11 kcal/mol contribution from core-valence connected quadruples.

Our W4.4 $D_0$, 116.22$\pm$0.10 kcal/mol at the CCSD(T)/cc-pV(Q+d)Z geometry and 116.27$\pm$0.10 kcal/mol at the CCSD(T)/cc-pwCVQZ geometry, at first sight seems significantly higher than the spectroscopic value of 116.06$\pm$0.09 kcal/mol. The latter (from a 1932 Herzberg paper\cite{Her32}) derives from a predissociation limit in the $C~^1\Sigma^+_u$ state to P($^4S$)+P($^2D$), where it is not clear whether the latter is in the $J=3/2$ or $J=5/2$ fine structure level. Taking the fine structure data from the NIST database, we obtain 116.10$\pm$0.07 kcal/mol in the former, and 116.06$\pm$0.07 in the latter case.  Especially the $J=3/2$ case agrees to within overlapping uncertainties with our calculations. 

We considered yet larger basis sets for calculating the post-CCSD(T) corrections, namely cc-pV5Z for $T_3-(T)$ and (Q), cc-pVQZ for $T_4-(Q)$, and cc-pVTZ for $T_5$. Results can be found in Table II. The higher-order triples contribution would change by -0.01 kcal/mol relative to W4.3 but by +0.01 kcal/mol relative to W4.4. The combined quadruples contributions would decrease by 0.04 kcal/mol, but in our experience\cite{pCb}, this basis set expansion tends to be mostly cancelled by the effect of expanding the basis set for the core-valence higher-order triples --- which, incidentally, is the reason (besides computational intractability) why all three expansions are neglected in W4.4 theory. Unfortunately, a CCSDT/cc-pwCVQZ calculation is not feasible (yet) for a molecule with two second-row atoms. Finally, the quintuples contribution would go up by 0.03 kcal/mol.

Let us now turn to P$_4$. The highest-level calculation we were able to perform was W4 theory, and even that stretched our computational resources (8-core AMD Opteron and Intel Cloverton machines with 16 to 32 GB of RAM and 2 TB of striped disk space each) to the limit. With a \%[(T)] diagnostic of 7.5\%, the CCSD(T) method is significantly in error here too; yet, we note 
that no single $T_2$ amplitude is larger than 0.05 for P$_4$.
We find a whopping -3.28 kcal/mol from higher-order triples, 2.57 kcal/mol from connected quadruples, and 0.13 kcal/mol from connected quintuples. 
The computed TAE$_0$ of 285.96$\pm$0.16 kcal/mol at the standard W4 reference geometry, and 286.09$\pm$0.16 kcal/mol at the core-valence reference geometry, agree well with the Gurvich TAE$_0$ of 285.9$\pm$1 kcal/mol, but this is a rather hollow victory in light of the stated uncertainty. We expect that the computed TAE$_0$ would go up significantly if we were able to perform W4.3 or W4.4 calculations. 

Interestingly, for P$_4$ CCSD(T) errs on the opposite side from the case of $D_0$(P$_2$), such that the error on the reaction energy P$_4\rightarrow 2$P$_2$ actually gets {\em amplified}: overall, CCSD(T) overestimates the full CI limit by 1.70 kcal/mol for this reaction. 

The experimental value actually derives from $D_0$(P$_2$) and the rather uncertain measured enthalpy for P$_4\rightarrow 2$P$_2$, evaluated in Gurvich et al.\cite{Gurvich} as 53.83$\pm$1 kcal/mol. As can be seen in Table I, our computed reaction energy for P$_4\rightarrow 2$P$_2$ converges fairly rapidly with the level of theory, and it can reasonably be assumed that the W4 value, 53.97 kcal/mol at the standard reference geometry and 54.00 kcal/mol at the core-valence geometry, is not far from Nature. Combining the latter with our W4.4 $D_0$(P$_2$) would suggest that W4 and W4.4 TAE$_0$ values would differ by about 0.4 kcal/mol, which is not impossible but does seem excessive. 286.53$\pm$0.10 kcal/mol would seem to be a plausible upper limit, and 286.10$\pm$0.16 a plausible lower limit. Intermediate values of 286.12 and 286.20 kcal/mol are obtained by combining the computed reaction energy with the two possible experimental $D_0$(P$_2$) values. Our best estimate, 286.32$\pm$0.22 kcal/mol, splits the difference between our estimated upper and lower limits.

Combining one-quarter of our best estimated TAE$_0$[P$_4$(g)] with one-quarter of the quite precisely known\cite{codata,Gurvich} $\Delta H^\circ_{f,0}$[P$_4$(g)]=15.83$\pm$0.07 kcal/mol, we obtain an estimated gas-phase heat of formation of phosphorus atom, $\Delta H^\circ_{f,0}$[P(g)]=75.54$\pm$0.1 kcal/mol. Using the CODATA heat content functions of P(g) and white phosphorus (see footnotes to Table I), we obtain $\Delta H^\circ_{f,298}$[P(g)]=75.74$\pm$0.1 kcal/mol, which is in excellent agreement with the CODATA reference value of 75.65$\pm$0.24 kcal/mol, but carries a much smaller uncertainty. Our calculations do seem to suggest that the true value is in the upper half of the CODATA uncertainty interval.

\section{Conclusions}

Summing up, the high-accuracy W4 computational thermochemistry protocol, and several post-W4 methods, have been applied to the P$_2$ and P$_4$ molecules. Contrary to previous studies, we find the experimental thermochemistry to be fundamentally sound. The quite significant contribution of post-CCSD(T) correlation effects to the reaction enthalpy for P$_4\rightarrow 2$P$_2$ illustrates the importance of their proper treatment in accurate computational thermochemistry work.
We derive a gas-phase heat of formation for the phosphorus atom of 75.54$\pm$0.1 kcal/mol, in the upper half of the CODATA uncertainty interval.

\acknowledgments

Research at Weizmann was funded by the Israel Science Foundation (grant
709/05), the Minerva Foundation (Munich, Germany), and the Helen and
Martin Kimmel Center for Molecular Design. JMLM is the incumbent of
the Baroness Thatcher Professorial Chair of Chemistry and a member
{\em ad personam} of the Lise Meitner-Minerva Center for Computational
Quantum Chemistry.  

The research presented in this paper is part of ongoing work
in the framework of a Task Group of the 
International Union 
of Pure and Applied Chemistry (IUPAC) on
'Selected free radicals and critical intermediates: thermodynamic
properties from theory and experiment'
(2000-013-2-100, renewal 2003-024-1-100).
See Ref.\cite{iupac1} for further details.


\clearpage
\squeezetable
\begin{table}
\caption{Component breakdown of total atomization energies of P$_2$ and P$_4$, as well as reverse dimerization reaction. Computed and CODATA heats of formation at absolute zero of phosphorus atom. All data in kcal/mol\label{tab:W4components}}
\begin{tabular}{lcc|c|c|c|c}
\hline\hline
 &&& P$_2\rightarrow$ & P$_4\rightarrow$ & P$_4\rightarrow$ &  \\
 &&&  2 P     &  4 P       &2P$_2$             & $\Delta H^\circ_{f,0}$[P(g)] \\
\hline
SCF            & aug-cc-pV(\{5,6\}+d)Z$^a$&W4 and W4.x  & 38.93 & 121.40 &  43.54 \\
CCSD           & aug-cc-pV(\{5,6\}+d)Z$^b$&W4,W4.2,W4.3 & 67.45 & 146.17 &  11.27 \\
                   & aug-cc-pV(\{5,6\}+d)Z$^{c}$& W4.4         & 67.43 & 146.12 &  11.27 \\
(T)            & aug-cc-pV(\{Q,5\}+d)Z$^d$& W4,W4.2,W4.3 &  9.71 &  21.82 &   2.40 \\
                   & aug-cc-pV(\{Q,5\}+d)Z$^{e}$& W4.4         &  9.67 &  21.70 &   2.36 \\
$\hat{T}_3$-(T)& cc-pV\{D,T\}Z$^f$ & W4,W4.2,W4.3 & -1.00 &  -3.28 &  -1.29\\
                   & cc-pV\{T,Q\}Z$^f$ & W4.3         & -0.94 & & & \\
                   & cc-pV\{T,Q\}Z$^{g}$ & W4.4         & -0.96 & & & \\
$\hat{T}_4$    & 1.1$\times$cc-pVTZ$^h$ & W4,W4.2      &  1.46 &   2.57 &  -0.34 \\
                   & cc-pV\{T,Q\}Z$^{i}$ & W4.3,W4.4    &  1.52 & & & \\
$\hat{T}_5$    & cc-pVDZ(no d) & W4,W4.2      &  0.10 &   0.13$^j$ &  -0.07 \\
                   & cc-pVDZ      &  W4.3,W4.4    &  0.13 & & & \\
$\hat{T}_6$        & cc-pVDZ(no d)  & W4.3,W4.4    &  0.03 & & & \\
inner              & CCSD(T)/aug-cc-pwCV\{T,Q\}Z &W4           &  0.73 &   1.76 &   0.31 \\
shell              &+$\Delta$CCSDT/cc-pwCVTZ& W4.2,3       &  0.76 & & & \\
                   &+(Q)/cc-pwCVTZ& W4.4         &  0.87 & & &\\
relativ.           &DK-CCSD(T)/aug-cc-pVQZ& W4 and W4.x  & -0.28 &  -0.72 &  -0.17 \\
DBOC$^{k}$           &HF/aug-cc-pVTZ & W4,W4.2,W4.3 &  0.01& 0.02 & 0.00 \\
                   &+CISD/cc-pVDZ & W4.4         &  0.01 & & & \\
A-M$^l$            && W4 only      &  0.02& 0.04& 0.00\\
ZPVE               &see text& W4           &  1.11& 3.91& -1.68\\
$\Delta E_{r,0}$   && W2.2         & 115.36& 286.60& 55.87& 75.61$^m$\\
                   && W3.2         & 115.41& 285.11& 54.28& 75.23\\
                   && W4lite       & 115.48& 285.03& 54.07& 75.22\\
                   && W4           & 116.00 & 285.96 & 53.97& 75.45\\
                   && W4.2         & 116.02 & & & \\
                   && W4.3         & 116.20 & & & [75.55]$^n$\\
                   && W4.4         & 116.22 & & & [75.56]$^n$\\
$\Delta$geom.      && all          & 0.05 & 0.13 & 0.03 & 0.03 \\
$\Delta E_{r,0}$   && best         & 116.27$\pm$0.10 & [286.32$\pm$0.22]$^o$ & & [75.54$\pm$0.1]$^n$\\
Experiment         && CODATA       & 116.0$\pm$0.7 & 285.9$\pm$1.0 & 53.87$\pm$0.7 & 75.45$\pm$0.24$^q$\\ 
                   && Gurvich      & 116.06$\pm$0.09$^p$ &          & 53.83$\pm$1 & 75.45$\pm$0.24$^q$\\ 
\hline\hline
\end{tabular}
\begin{flushleft}
The notation cc-pV\{X,Y\}Z, for instance, indicates extrapolation from cc-pVXZ and cc-pVYZ basis sets.\\
$^a$ extrapolated using the Karton-Martin\cite{MKtca} modification of Jensen's extrapolation formula\cite{JensenExtrap}\\
$^b$ $\hat{T}_1$ term from largest basis set, singlet- and triplet-coupled pair energies are extrapolated using the two-point $A+B/L^\alpha$ expression of Halkier et al.\cite{Hal98} with $\alpha=$3 and 5, respectively.\\
$^{c}$ same expressions, but using Schwenke's extrapolation\cite{Schwenke2005}, which is equivalent to $\alpha=$3.06967 for singlet pairs and $\alpha=$4.62528 for triplet pairs.\\
$^d$ extrapolated using the two-point $A+B/L^3$ expression of Halkier et al.\cite{Hal98}\\
~\\
(Footnotes continued on next page of manuscript.)
\end{flushleft}
\end{table}
\clearpage
{\scriptsize
\begin{flushleft}
$^{e}$ same expression, but using Schwenke's extrapolation\cite{Schwenke2005}, which is equivalent to $\alpha=$3.22788\\
$^f$ extrapolated from AVQZ and AV5Z basis sets using the two-point $A+B/L^3$ expression of Halkier et al.\cite{Hal98}\\
$^{g}$ extrapolated using empirical $A+B/L^{5/2}$ found in Ref.\cite{pCb}\\
$^h$ Approximated as 1.1$\times$(Q)/cc-pVTZ~+~1.1$\times$[Q-(Q)]/cc-pVDZ, as proposed in Ref.\cite{w4}\\
$^{i}$ (Q) contribution extrapolated cc-pV\{T,Q\}Z, Q-(Q) contribution obtained as difference between CCSDTQ and CCSDT(Q) calculations with the cc-pVTZ basis set.\\
$^j$ Value for P$_4$ is approximated by CCSDTQ5$_{\Lambda}-$CCSDTQ\\
$^{k}$ Post-Hartree-Fock corrections to the DBOC as obtained at the CISD/cc-pVDZ level are found to be +0.003 kcal/mol for P$_2$ and +0.001 kcal/mol for P$_4$. This correction exhibits very weak basis set sensitivity: A. Karton, B. Ruscic, and J. M. L. Martin, 
{\em J. Mol. Struct. ({\sc theochem})} {\bf 811}, 345-353 (2007).\\
$^l$ differences between ACES II and MOLPRO definitions of the valence CCSD(T), extrapolated from PVTZ and PVQZ basis sets using the two-point $A+B/L^3$ expression of Halkier et al.\cite{Hal98}\\
$^m$ Computed $\Delta H^\circ_{f,0}$[P(g)] values are derived as (1/4)($\Delta H^\circ_{f,0}$[P$_4$(g)]+TAE$_0$[P$_4$(g)]), where the experimental $\Delta H^\circ_{f,0}$[P$_4$(g)]=15.83$\pm$0.07 kcal/mol is obtained from CODATA $\Delta H^\circ_{f,298}$[P$_4$(g)]=58.9$\pm$0.3 kJ/mol, [$H_{298}-H_0$][P$_4$(g)]=14.10$\pm$0.20 kJ/mol, and [$H_{298}-H_0$][P(cr,white)]=5.360$\pm$0.015 kJ/mol. Note that using the best currently available fundamental frequencies\cite{Kor02} for P$_4$, the rigid rotor-harmonic oscillator [$H_{298}-H_0$][P$_4$(g)]=14.046 kJ/mol.\\
$^n$ Assuming that $\Delta E_{r,0}$ for the dimerization reaction, 2~P$_2\rightarrow$P$_4$, is converged at the W4 level. Further basis set effects then amount to $(1/2)\Delta D_0$(P$_2$).\\
$^o$ See the text.\\
$^p$ Spectroscopic value extracted from Herzberg\cite{Her32}.
Value derives from predissociation of $C~^1\Sigma^+_u$ state to P($^4S_{3/2}$)+P($^2D_x$), where $x\in \{3/2,5/2\}$. Combining Herzberg's predissociation limit of 51969$\pm$24 cm$^{-1}$ with the latest atomic excitation energy data from the NIST database\cite{nist} ($^2D_{3/2}$: 11361.02 cm$^{-1}$ and $^2D_{5/2}$ 11376.63 cm$^{-1}$), we obtain 
116.10$\pm$0.07 kcal/mol if $^2D_{3/2}$ and 
116.06$\pm$0.07 kcal/mol if $^2D_{5/2}$. \\
$^q$ From CODATA $\Delta H^\circ_{f,298}$[P(g)]=316.5$\pm$1 kJ/mol, 
[$H_{298}-H_0$][P(g)]=6.197$\pm$0.001 kJ/mol, and 
[$H_{298}-H_0$][P(cr,white)]=5.360$\pm$0.015 kJ/mol.
\end{flushleft}
}
\clearpage

\squeezetable
\begin{table}
\caption{Basis set convergence of various contributions to the total atomization energy (kcal/mol)}
\begin{tabular}{l|cccc|cccc}
\hline\hline
& \multicolumn{4}{c}{P$_2$}&\multicolumn{4}{c}{P$_4$}\\
\hline
& AV(T+d)Z & AV(Q+d)Z & AV(5+d)Z & AV(6+d)Z& AV(T+d)Z & AV(Q+d)Z & AV(5+d)Z & AV(6+d)Z\\
\hline
SCF &  38.35 & 38.72 & 38.83 & 38.91 & 120.24 & 120.83 & 121.17 & 121.36 \\
CCSD T-pair & 23.86 & 24.45 & 24.58 & 24.61 & 66.64 & 68.94 & 69.50 & 69.66 \\
CCSD S-pair & 36.07 & 40.27 & 41.73 & 42.32 & 59.95 & 69.79 & 73.54 & 75.01 \\
CCSD $T_1$  & -0.29 & -0.30 & -0.31 & -0.31 & -0.58 & -0.60 & -0.62 & -0.62 \\
CCSD total  & 59.65 & 64.42 & 66.01 & 66.62 & 126.01 & 138.13 & 142.43 & 144.05 \\
(T)         & 9.09 & 9.36 & 9.53 & 9.59 & 19.79 & 20.82 & 21.31 & ---\\
\hline
& PVDZ & PVTZ & PVQZ & PV5Z & PVDZ & PVTZ & PVQZ & PV5Z \\
\hline
$T_3$-(T)   &-0.07 & -0.72 & -0.85 & -0.90 & -0.78 & -2.54 & ---& ---\\
(Q)         & 1.04  & 1.43 & 1.57 & 1.61 & 1.77 & 2.66 & ---& ---\\
$T_4$-(Q)   & -0.11 & -0.15 & -0.17 & --- & -0.32 & --- & --- & --- \\
$T_5$       & 0.13$^a$ & 0.16 & --- & --- & (a) & --- & --- & --- \\
\hline
& ACVTZ & ACVQZ &      &       & ACVTZ & ACVQZ \\
\hline
core-valence & 0.67 & 0.70 & & & 2.00 & 1.86 \\
\hline\hline
\end{tabular}
\vspace*{12pt}
\begin{flushleft}

(a) With PVDZ(no d) basis set: 0.10 kcal/mol for P$_2$, 0.13 kcal/mol for P$_4$.
\end{flushleft}

\vspace*{1in}

\end{table}

\squeezetable
\begin{table}
\caption{Diagnostics for importance of nondynamical correlation\label{tab:diagnostics}}
\begin{tabular}{l|cccc|ccccc}
\hline\hline
 & \%TAE[SCF]$^a$ & \%TAE[(T)]$^a$ & \%TAE & \%TAE[$T_4+T_5$]$^a$ & ${\cal T}_1$\cite{Lee89t1} & $D_1$\cite{D1diagnostic} & Largest T$_2$ &   \multicolumn{2}{c}{NO occupations}  \\
 &  & & [post-CCSD(T)]$^a$ &  & \multicolumn{2}{c}{diagnostic} & amplitudes  & HDOMO$^b$ & LUMO  \\
 &  &  &  &  & \multicolumn{5}{c}{~~~~~~~~--- CCSD(T)/cc-pVTZ ---}   \\
\hline
P$_2$     & 33.2 & 8.29 & 0.48 & 1.33 & 0.018 & 0.033 & 0.118 (x2) & 1.899 & 0.085 \\  
P$_4$     & 41.9 & 7.53 &-0.20 & 0.93 & 0.018 & 0.037 & 0.048      & 1.916 & 0.061 \\  
\hline\hline
\end{tabular}
\begin{flushleft}
$^a$Percentages of the total atomization energy relate to nonrelativistic, clamped-nuclei values with inner shell electrons constrained to be doubly occupied. (from W4 theory)\\
$^b$ Highest doubly occupied molecular orbital.\\

\end{flushleft}
\end{table}

\end{document}